# Deep Reinforcement Learning-based Cell DTX/DRX Configuration for Network Energy Saving

Wei Mao
*Intel Corporation*
Santa Clara, United States
wei.mao@intel.com

Lili Wei
*Intel Corporation*
Santa Clara, United States
lili.wei@intel.com

Omid Semiari
*Intel Corporation*
Santa Clara, United States
omid.semiari@intel.com

Shu-ping Yeh
*Intel Corporation*
Santa Clara, United States
shu-ping.yeh@intel.com

Hosein Nikopour
*Intel Corporation*
Santa Clara, United States
hosein.nikopour@intel.com

*Abstract*—3GPP Release 18 cell discontinuous transmission and reception (cell DTX/DRX) is an important new network energy saving feature for 5G. As a time-domain technique, it periodically aggregates the user data transmissions in a given duration of time when the traffic load is not heavy, so that the remaining time can be kept silent and advanced sleep modes (ASM) can be enabled to shut down more radio components and save more energy for the cell. However, inevitably the packet delay is increased, as during the silent period no transmission is allowed. In this paper we study how to configure cell DTX/DRX to optimally balance energy saving and packet delay, so that for delay-sensitive traffic maximum energy saving can be achieved while the degradation of quality of service (QoS) is minimized. As the optimal configuration can be different for different network and traffic conditions, the problem is complex and we resort to deep reinforcement learning (DRL) framework to train an AI agent to solve it. Through careful design of 1) the learning algorithm, which implements a deep Q-network (DQN) on a contextual bandit (CB) model, and 2) the reward function, which utilizes a smooth approximation of a theoretically optimal but discontinuous reward function, we are able to train an AI agent that always tries to select the best possible Cell DTX/DRX configuration under any network and traffic conditions. Simulation results show that compared to the case when cell DTX/DRX is not used, our agent can achieve up to ~45% energy saving depending on the traffic load scenario, while always maintaining no more than ~1% QoS degradation.

## I. INTRODUCTION

Energy saving in radio access network (RAN) is critical to reducing network operating cost and following environmental stringent requirements, while ensuring service level agreement in a cellular network. According to recent industrial whitepapers [1] [2], RAN accounts for ~73% of the total operator energy use [1], and a 15% saving in RAN energy would save $165M annually for a major network operator [2]. In addition, in remote areas where the energy resources are limited, energy saving of network nodes ensures the continuity of the service delivery under the limited energy resource constraints. To enable RAN network energy saving (NES), 3GPP defines four major categories of techniques in [3] including frequency, time, spatial and power domain approaches, to make more efficient use of radio resources, particularly in low/medium load scenarios.

In this paper we focus on the cell discontinuous transmission and reception (DTX/DRX) feature introduced in 3GPP Release 18 (Rel-18), which is a layer 2 (L2) time-domain technique to enable advanced sleep mode (ASM) and radio unit (RU) shutdown. It is an effective energy saving technique since among the equipment of a typical base station (BS), about 80% of the energy is consumed by the RU [4]. Cell DTX/DRX operates by enabling the medium access control (MAC) layer to pack low to moderate traffic into a smaller number of transmission time intervals (TTI), while the rest of the TTIs remain silent. During such inactive time, the BS can enter a level of ASM and shut down some components of RU. The longer the inactive time is, the more RU components the BS can shut down, and so the greater the energy saving is. Deeper level of ASM, however, comes at the expense of higher transmission latency as packets have to wait longer for the BS to reactivate its RU components and start the transmission or reception again. Hence it is of great interest to study how to configure cell DTX/DRX optimally to maximize energy saving while ensuring quality of service (QoS), which is the problem we try to solve in this work.

As a new feature, 3GPP Rel-18 cell DTX/DRX has not yet received enough studies in the literature, especially on its configuration optimization. The general concepts of DTX/DRX, though, exist since LTE and mainly include cell DTX [5] and user equipment (UE) DRX [6]. The early work [5] shows the great energy saving potential of cell DTX in LTE (Release 8) systems and possible ways to further enhance it. To optimize its performance [7] and [8] consider using reinforcement learning (RL) methods [9] together with fuzzy logic [7]. UE DRX is defined in both LTE and 5G new radio (NR), with ample literature [6]. Among them we find [10] and [11] are most relevant to our problem, where [10] tries to employ RL methods such as contextual bandit (CB) to find an optimal UE DRX configuration that achieves the optimal balance between UE throughput and energy consumption. [11], on the other hand, utilizes deep reinforcement learning (DRL) methods to jointly optimize the UE DRX and bandwidth part (BWP).

None of the above literature is directly applicable to our cell DTX/DRX optimization problem, since its configuration as defined in 3GPP Rel-18 [3] [12] is different from either LTE/5G UE DRX or LTE cell DTX. In this paper, we leverage the open RAN (O-RAN) [13] architecture and utilize RL methods including contextual bandits (CB) and deep Q-network (DQN) [9] to learn an artificial intelligence (AI) agent that tries to always select the best cell DTX/DRX configuration under different network conditions and different traffic loads. The standard open interfaces in O-RAN facilitate data collection from the RAN to enable training of the agent, which is deployed as an xApp on the near-real time RAN intelligent controller (Near-RT RIC) [13]. The agent monitors the cell traffic conditions and data transmission status through the E2 interface [13] to infer an optimized selection of cell DTX/DRX configuration parameters, which is also updated periodically to reflect the changes in the network conditions and traffic load profile.

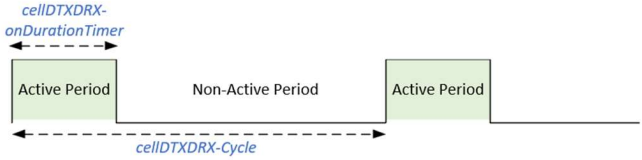

Fig. 1. Cell DTX/DRX illustration.

The remainder of the paper is organized as follows. Section II briefly describes 3GPP Rel-18 cell DTX/DRX configuration and the power consumption model used in the paper, and then formulates our cell DTX/DRX optimization problem. Section III provides the details of our DRL-based solution, whose performance evaluation is described in Section IV, using 5G system-level simulations (SLS). Section V concludes the paper.

## II. SYSTEM MODEL AND PROBLEM FORMULATION

### A. Description of Rel-18 Cell DTX/DRX

Cell DTX/DRX mechanism is introduced in 3GPP Rel-18 as an important NES feature. As shown in Fig. 1, in time domain the cell operation is divided into periodic DTX/DRX cycles and within each cycle, there are cell transmission/reception active period and non-active period. The cell operates normally during an active period, but there is no data transmission/reception together with corresponding layer 1 (L1) signals/channels during a non-active period, allowing the activation of ASM.[1] Cell DTX/DRX typically operates at a relatively small time scale, e.g., from symbol-level to hundreds of milliseconds (ms) and up [12]. It provides an opportunity for time-domain energy saving if completely switching off a cell is not possible (which operates at a relatively large time scale, e.g., on the order of at least tens of minutes.) It has been observed that cell DTX/DRX can achieve 47.8%~71.3% energy saving gain under different system loads [14], with more NES gain achieved when the system load decreases.

3GPP signaling mechanisms for cell DTX/DRX operation include radio resource control (RRC) and L1 downlink control information (DCI) signaling. The cell DTX/DRX RRC parameters include [12] configuration type, (initial) activation status, cycle length (i.e., periodicity), on-duration, start offset (for a cycle) and slot offset (for the active period), whose possible values are listed in Table I (extracted from Sec 6.3.2, [12]). The L1 group signaling, in particular DCI format 2_9 [15], can be used to quickly enable or disable the RRC-configured cell DTX/DRX pattern, which is useful when urgent delay-sensitive transmission is needed.

---

[1] Note that during cell DTX/DRX non-active period, the cell continues to support synchronization signal block (SSB) transmission, random access procedure, paging and system information broadcast, etc., to maintain basic cell operations. However, 3GPP allows longer periodicity of such signaling transmissions (e.g., up to 160ms [16]) to facilitate NES. Discussions of on-demand SSB in some scenarios are also on-going in 3GPP for Release 19. In this work, as an initial study of this topic, we ignore the impact of such signaling transmissions for simplicity (in Section IV we will see that the periodicity we considered is much less than 160ms, which justifies our simplified assumption to some extent).

TABLE I. CELL DTX/DRX RRC PARAMETER VALUES

| Parameter name | Possible values |
|---|---|
| Configuration type | DTX, DRX, or both |
| Activation status | Activated, deactivated |
| Cycle length | {10, 20, 32, 40, 60, 64, 70, 80, 128, …,10240} ms |
| On-duration timer | {1,…,31}/32 & {1, 2, 3, 4, 5, 6, 8, 10, 20, 30, 40, 50, 60, 80, 100, …, 1600} ms |
| Start offset | Any integer ms smaller than cycle length |
| Slot offset | {0,1,…,31}/32 ms |

TABLE II. EXAMPLE OF POWER CONSUMPTION AND TRANSITION TIME AND ENERGY

| Power state | Characteristics | Relative Power | Trans. time | Trans. energy |
|---|---|---|---|---|
| Deep sleep (SM3) | Most of PHY blocks are turned off. | 1 | 50 ms | 1000 |
| Light sleep (SM2) | Many PHY blocks are turned off. | 25 | 6 ms | 90 |
| Micro sleep (SM1) | Some RF components are turned off. | 50 | 0 | 0 |
| Active DL | DL transmission. | 200 | N/A | N/A |
| Active UL | UL reception. | 90 | N/A | N/A |

In this paper we only focus on the RRC signaling and leave L1 group signaling for future study. For simplicity we only study the downlink (DL) side of the system and assume the configuration type is DTX, since the cell is in transmission mode for DL and its power consumption dominates the uplink (UL) case, when the cell is in reception mode. Furthermore, we assume the TTI duration of the system is 1ms, again for simplicity, and thus ignore the fractional values of on-duration, as well as the slot offset parameter (i.e., setting to 0).

### B. Sleeping Modes and Power Consumption Model

Depending on how long the cell is inactive (for example, during cell DTX/DRX non-active period), the cell can shut off/deactivate certain parts, or even most parts of its hardware components to enable different sleep modes (SMs) and save energy. 3GPP defines three different levels of SMs [3] and an example of the relative power consumption, as well as the total transition (entering and leaving) time needed, together with the associated relative energy consumption (expressed in relative power * ms) for each sleeping mode are shown in Table II (extracted from Table I and III of [16], using BS configuration set 2). Note that in this table, the active DL and UL power values are the full power when all system bandwidth (BW) is used. If only a fraction $s_f$ of BW is used, for DL relative power we use the following simplified scaling expression:[2]

$$P_{DL}(s_f) = 110 + 90s_f. \quad (1)$$

Moreover, for the deep and light sleep modes (SM3 and SM2), 3GPP further requires the time duration for the sleep to be larger than the total transition time entering and leaving this state [3].

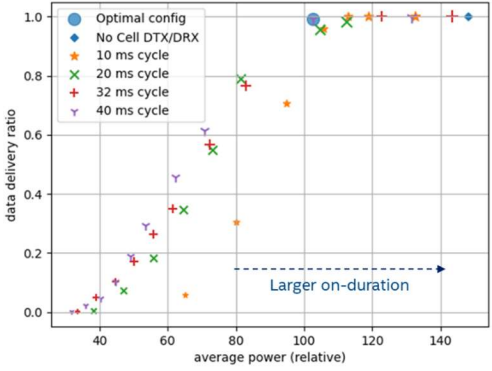

Fig. 2. Optimal cell DTX/DRX configuration.

*C. Problem Formulation*

Since the packets that arrive during the non-active period need to wait until the next active period before they can be transmitted, cell DTX/DRX increases packet delay and this delay grows in proportion to the duration of the non-active period. Thus, for delay-sensitive traffic, there is a trade-off between achieving the QoS and saving more energy: given the traffic intensity, if the cycle length is too long or the on-duration is too short, then within the cell DTX/DRX active period there might not be enough time to timely deliver all packets that arrive within a cycle. On the other hand, if the cycle length is too short or the on-duration is too long, then the cell cannot enable ASM, or the sleep duration is too short, which reduces energy saving. As shown in Fig. 2, for a given traffic load distribution on a cell, there is an optimal choice of the cell DTX/DRX parameters *(cycle length, on-duration)* that enables maximum energy saving while maintaining the QoS (that means, in the case of Fig. 2, delivering all packets within their required latency so that there are no packet drop due to delay violation). For different cells with different traffic load conditions (and different delay requirements), the optimal cell DTX/DRX configuration may be different, and finding that configuration under each traffic scenario for each cell is a complex problem.

## III. DRL-BASED CELL DTX/DRX CONFIGURATION

In this paper we utilize the DRL framework to train an AI agent to solve this configuration optimization problem, aiming to select the best possible cell DTX/DRX configuration for every network and traffic scenario. In Fig. 3, we illustrate our DRL-based solution as an xApp in the O-RAN architecture. Similar to other ML/AI-based xApps, its operation is divided into training and inference modes, and its interaction with the RAN environment is through the E2 interface. Periodically for each cell, the xApp observes RAN measurements and conditions to make a decision on the DTX/DRX RRC configuration for the next period. If it is in the training mode, in each period it also collects some performance and power metrics data from RAN to form an RL reward to update its AI agent. Such periods are called observation periods, which are also the RL steps. They can have different durations for the training and inference modes, but in both cases they should be long enough (covering multiple DTX/DRX cycles, e.g., ~1 second (s)) to enable reliable RAN observation and/or metrics data collection.

<sup>2</sup> Obtained from (1)-(3) of [16] by setting $s_a = s_p = \eta = 1$.

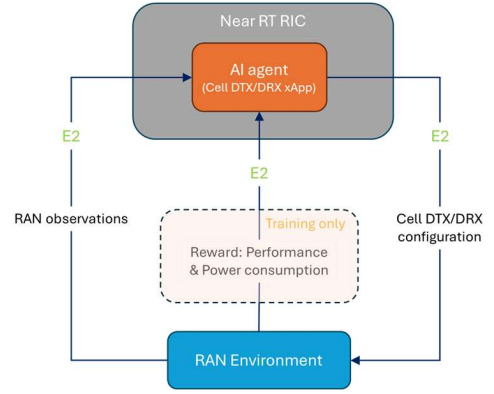

Fig. 3. Cell DTX/DRX xApp in ORAN architecture.

TABLE III. RAN OBSERVATIONS (FROM EACH CELL)

| Category | Observation | Measurement(s) |
|---|---|---|
| 1 | Traffic intensity | total amount of data that arrived during the observation period, divided by the period duration |
| 1 | Inter-arrival time statistics | mean and variance of the time between each two adjacent packet arrivals, during the observation period |
| 1 | Packet size statistics | mean and variance of the sizes of the packets that arrived during the observation period |
| 1 | Traffic delay requirements | minimum and the (weighted) average delay requirements of the packets that arrived during the observation period |
| 2 | Cell transmission capability | amount of transmitted data divided by the proportion of radio resource used for each TTI, averaged over the observation period |

The input to the xApp are the RAN observations from the cells, and in training mode, the RAN metrics data for forming rewards as well. Our selection of RAN observations includes two categories of information: 1) traffic profile, and 2) RAN transmission conditions. The detailed observations in the two categories and how to measure them are listed in Table III. Note that with our simplifying assumptions in Section II, the radio resource proportion for each TTI can be measured by the physical resource block (PRB) utilization, which is also the BW fraction $s_f$ in (1).

The RAN metrics data for reward formation include the QoS performance and power metrics. For the QoS performance metric we use the delivered data ratio $y$, which is defined as follows. Out of all packets that arrive within the observation period, let $y_d$ be the amount of data contained in the timely received packets, and let $y_f$ be that contained in the packets that are not successfully received in time. Then we define the delivered data ratio as

$$y \coloneqq y_d/(y_d + y_f). \quad (2)$$

For the power metric we use the normalized power consumption $x$, defined as the ratio of the average cell power consumption $x_a$ of the observation period, to the maximum possible cell power consumption $x_m$:

$$x \coloneqq x_a/x_m. \quad (3)$$

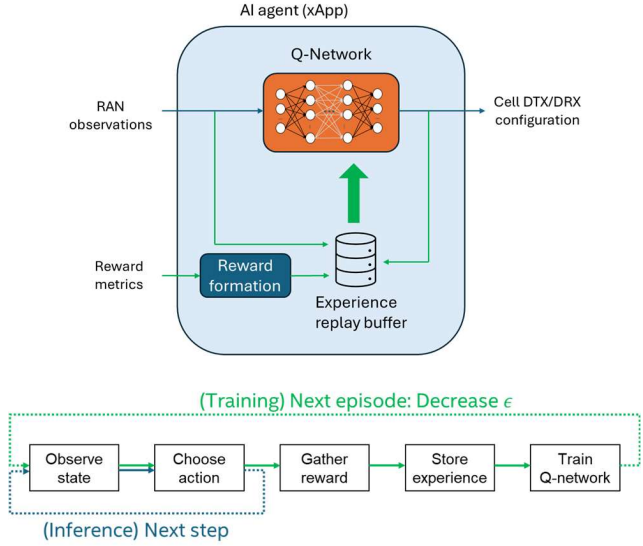

Fig. 4. Operation of the cell DTX/DRX xApp. Upper: agent design; lower: simplified algorithm flow. Green arrows are for training mode only.

Note that if 3GPP power model is used as in Section II.B, then all power quantities in the formula are relative. In particular, $x_m = 200$ for DL power.

The output of the xApp is the cell DTX/DRX RRC configuration. It mainly sets the following parameters for each cell: *cycle length*, *on-duration*, and *start offset*. (Note that how other parameters are set in this work is described in Section II.A.) In the following we describe how we design and train our AI agent to learn the optimal configuration of these parameters, especially the first two, based on the input data.

### A. DRL Algorithm Design: DQN with CB

The design and operation of our AI agent is illustrated in Fig. 4. We use a CB-based agent, which treats the RAN observations of each cell as a state (context), and for each possible action (representing one particular cell DTX/DRX RRC configuration) at this state it associates an expected reward. The DQN algorithm is then used to train the CB agent, which uses a Q-network [9] to model the reward function, together with an experience replay buffer to store the history of *(state, action, reward)* tuples, which are called experiences.

The reason why we choose CB over an ordinary Markov decision process (MDP) as the underlying RL model is as follows. As stated earlier, the observation period/duration of RL step should be long enough to enable reliable RAN observation and/or metrics data collection. In practice we observe that due to the randomness of the measurements (similar phenomenon is also reported in [10]), the observation period should be at least 1 s or larger to differentiate the performances of different actions at the same state. For step durations of this length, the correlations between the states (RAN observations) of consecutive steps are quite weak and it is reasonable to assume that they are approximately independent. In this case, the ordinary multi-step MDP decomposes to multiple episodes of 1-step MDP, which is exactly CB. This treatment greatly reduces the training time on SLS, since instead of simulating multiple steps of the same episode (network deployment), we can now simulate more independent episodes with only 1-step, which introduces more independence in the collected data (experiences) and is known to be beneficial to the Q-network training [9].

The Q-network we use is a fully-connected multi-layer neural network (similar to the architecture used in [17]), which takes the RAN observations (state) of each cell as input and outputs the predicted reward for each action under the input state. Thus the number of nodes in the input layer is the same as the number of RAN observation measurements in Table III, while the number of nodes in the output layer is the number of allowed actions, each output node corresponding to exactly one action. These actions are the cell DTX/DRX RRC parameter pairs *(cycle length, on-duration)*, which may take any of the finite values in Table I, provided that 1) the cycle length is smaller than the minimum latency requirement of all traffic (provided as a system parameter), and 2) the on-duration is smaller than the cycle length. In addition, we add the pair (1,1) to the action space to represent the configuration that sets the system to be always active, i.e., when the cycle length and on-duration are equal.

Between the input and output layers we use two hidden layers of the same size 128, whose nodes form linear combinations of the previous layer output and apply a ReLU (rectified linear unit) activation function to them, which converts any negative results to zero. The output layer forms a linear combination of the last hidden layer output at each output node, as the predicted reward for the corresponding action (at the input state). Furthermore, to speed up training, each input needs to go through a normalization process before sending to the next layer.

The experience replay buffer stores up to a given number of (for example, 10000) most recent experience entries for training. As new experiences come in, the oldest experiences are removed from the buffer. In the training mode, the RAN observations (states), reward, and DTX/DRX RRC configuration (action) for each observation period are collected per cell to form the new experience entries ((i.e., the *(state, action, reward)* tuples) and are stored in the experience replay buffer (see Fig. 4). Once the number of experience entries reaches a given threshold (e.g., 10 times of the batch size (defined blow)), for each RAN observation measurement the maximum value across these entries is calculated to serve as the normalizer of the corresponding input node of the Q-network.

Then the training starts, with one or more training step being performed for each observation period (i.e., each episode). In our experiments in Section IV, we set the number of training steps to be the same as the number of experiences collected for each observation period. For each training step, a batch of experience entries of given size (e.g., 128) is randomly drawn from the experience replay buffer. For each entry in the batch, the state and action fields are provided to the Q-network to obtain a predicted reward, which is then compared to the actual reward in the experience entry to produce a reward difference. The reward differences created from all entries of the batch are then used to perform an optimization step of the Q-network, using a smooth L1 loss function [17] with a chosen optimization algorithm, such as the Adam optimization algorithm [18].

To decide what DTX/DRX RRC configuration (action) to use in each observation period in the training mode, we use the $\epsilon$-greedy algorithm [9], where $\epsilon$ is an exploration probability that decreases gradually from a larger value (e.g., 0.9) to a small value (e.g., 0.05). With probability $\epsilon$, an action is selected randomly from all allowed actions (called *exploration*). With probability $1 - \epsilon$, the action is formed as follows (called *exploitation*): for each cell, the current state is fed to the Q-network, which outputs a predicted reward on each output node. The action that corresponds to the node producing the largest predicted reward is then chosen to be the action for the cell, which determines the cycle length and on-duration for the DTX/DRX configuration. In addition, the start offset is set to zero to reduce measurement randomness for better training.

The training finishes after the convergence of Q-values (i.e., the predicted rewards) is observed. Then the agent is ready to be deployed for inferencing. In the inferencing mode, for each cell, the RAN observations are fed to the Q-network. Then the cycle length and on-duration for the DTX/DRX configuration are determined in the same way as the exploitation case of the $\epsilon$-greedy algorithm above. The start offset is set differently, though: in inference mode, we set it to a random value among those allowed in Table I. The purpose of this setting is to reduce the inter-cell interference (ICI): when all cells have the same start offset value, their DTX/DRX cycles are aligned and they start to transmit at the same time, since a cycle always starts with an active period (see Fig. 1, ignoring slot offset). As a result, the interference level is elevated. By setting a random start offset value, the neighboring cells can have staggered active periods and less concurrent transmissions, thus reducing ICI.

### B. Reward Function Design

As is the case with any RL algorithm, the design of an appropriate reward function is crucial to making sure our learning objective is achieved. For the cell DTX/DRX optimization problem, the objective is to find the optimal balance between the QoS performance and power consumption. For each observation period, the former is measured by the delivered data ratio $y$, whereas the latter is measured by the normalized power consumption $x$, as defined in (2) and (3), respectively.

The most straightforward way to design a reward that balances these two metrics is to form a linear combination of them. Hence we define our first reward function, the *linear reward* as

$$r_{lin} = -(1 - c)x - c(1 - y), \quad (4)$$

where $1 - y$ represents the failed data ratio and $0 \leq c \leq 1$ is the coefficient that controls the trade-off between QoS performance and energy saving. We have negative signs for both terms because we want a higher reward when there is more energy saving (smaller $x$) or higher QoS (smaller $1 - y$).

The linear reward is simple and amenable to RL training, since it is continuous and has a constant gradient. However, the balancing coefficient $c$ needs to be carefully selected to guide the AI agent towards the best action (i.e., the optimal configuration in Fig. 2). Furthermore, for different network and traffic conditions, the choices of $c$ may need to be different, or even conflicting. Therefore, we consider a more principled design of the reward function. If our QoS metric $y$ is required to be no smaller than some threshold $y_0$ (i.e., the target QoS), then when the current metric $y < y_0$, the agent should be guided towards a larger $y$ without considering $x$, since we do not want to consider energy saving if QoS is not satisfied yet. When $y \geq y_0$, however, the agent should be guided towards a larger energy saving (i.e., smaller $x$) without considering $y$, since the QoS is already satisfied. This leads to our definition of the *QoS-threshold reward*

$$r_{QoS} = \begin{cases} -ax & \text{if } y \geq y_0 \\ y - b & \text{o.w.} \end{cases}, \quad (5)$$

where the constants $a > 0$ and $b$ satisfies $-a \geq y_0 - b$, since we always want to guarantee the QoS first: that means, the rewards in the region $y \geq y_0$ is never smaller than the region $y < y_0$. One example of these constants is $a = 1, \ b = 1 + y_0$.

In theory, the QoS-threshold reward is optimal and can precisely guide the AI agent towards the best action (e.g., when applied to Fig. 2 with $y_0$ close to 1). However, there is a discontinuity at the region boundary $y = y_0$, especially for smaller $x$ (since $0 \leq x \leq 1$ and $-ax \geq -a \geq y_0 - b$). This creates a big training stability issue, due to the aforementioned randomness of the RAN measurements (including the metrics $x$ and $y$). Assume the performance metrics $(x, y)$ of an action is close to the boundary $y = y_0$. Then because of this randomness, the measured $(x, y)$ could land on either side of the boundary, giving very different rewards (5), which makes the training very unstable. To address this issue, we seek a smooth approximation of (5). Our choice of approximation function is:

$$r_{QoS\_approx} = -\frac{u[1 + (\alpha - 1)(1 - y)] + x}{u + 1}, \quad (6)$$

where $u \coloneqq \left[\frac{1-y}{(1-y_0)(1-x)}\right]^m$. We call this reward function the *approximated QoS reward*. The constant $0 < y_0 < 1$ resembles the QoS threshold above but is only approximate, as it is "modulated" by the energy saving term $1 - x$. The constants $m > 0$ and $\alpha > 1$. An example set of these constant values is $y_0 = 0.9, m = 2,$ and $\alpha = 2$.

Assume $m$ is large. When $1 - y > (1 - y_0)(1 - x)$, resembling the case that the QoS is not satisfied yet (since the failed data ratio $1 - y$ is relatively large), the ratio in the definition of $u$ is larger than 1 and so $u \gg 1 \geq x$, since $m$ is large. Then $r_{QoS\_approx} \approx -[1 + (\alpha - 1)(1 - y)]$, which only grows with $y$, and so it will guide the agent towards better QoS only. On the other hand, when $1 - y < (1 - y_0)(1 - x)$, resembling the case that the QoS is already satisfied, the ratio in $u$ is less than 1 and so $u \approx 0$, again as $m$ is large. In this case, $r_{QoS\_approx} \approx -x$, which only grows if $x$ decreases, and so the agent will be guided toward a larger energy saving only. The "modulation" term $1 - x$ allows a larger comparison threshold $(1 - y_0)(1 - x)$ to be applied when $x$ is small, which provides more room for transition on the boundary when the difference between the two sides are larger (i.e., cf. (5), when $x$ is smaller), and thus allows for a smoother approximation.

The constant $m$ controls the trade-off between the smoothness of the reward function and the accuracy of the approximation. When $m$ is larger (as described above), the approximation is more accurate but the reward function is less smooth (since $r_{QoS}$ is discontinuous). When $m$ is smaller, however, the approximation is less accurate but the reward is smoother and better for training. The constant $\alpha$ controls the maximum amplitude of the reward (minimum reward in fact, since the reward is always negative). The reward $r_{QoS\_approx}$ is smooth and thus much more stable than $r_{QoS}$, while in the meantime, it tries to approximate the surface shape of the latter. Thus in training it can also provide a good guidance to the agent, as is evidenced by the experimental results in the next section.

TABLE IV. SIMULATION PARAMETERS

| Parameter | Value |
|---|---|
| Carrier frequency, channel BW | 725 MHz, 5 MHz |
| Subcarrier spacing | 15 kHz |
| TTI duration | 1 ms |
| Cell transmit power | 49 dBm |
| Packet size | {125, 150, ..., 500} bytes |
| Mean inter-arrival time | {10, 15, 20} ms |
| Delay requirement | {50, 75, 100} ms |

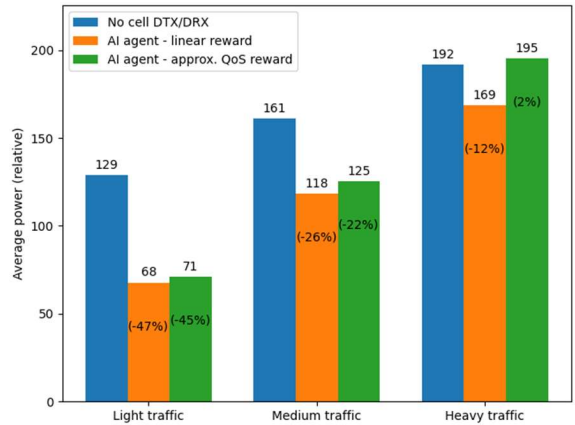

Fig. 6. Power consumptions of AI agents and baseline.

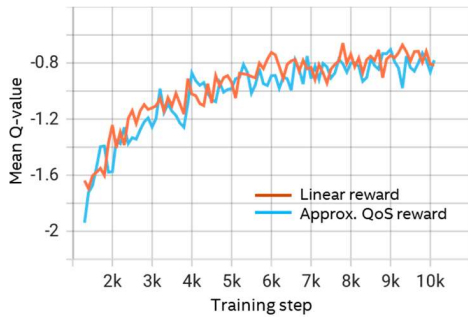

Fig. 5. Convergence of mean Q-values (sampled every 100 steps).

## IV. PERFORMANCE EVALUATION

To evaluate the performance of our proposed DRL solution we use a detailed 5G Python-based system-level simulator (SLS), supporting 3GPP compliant channel models, non-ideal CSI feedback, non-full buffer traffic, link adaptation, RLC re-transmissions, and PF scheduling. The network is composed of 7 BS sites randomly deployed in a 1km × 1km simulation area, with a minimum inter-site distance of 35m. Each site includes 3 sectors, resulting in a total of 7 × 3 = 21 cells. For each deployment, 210 UEs (10 UEs per cell on average) are distributed randomly within the simulation area. The UE traffic model is adapted from FTP model 3 [19]. In particular, for each UE the packet size, mean inter-arrival time, and delay requirement are all randomly selected from their respective ranges at the beginning of the simulation. These ranges and some of the other main simulation parameters related to the wireless system are listed in Table IV.

Our DRL algorithm is implemented in PyTorch and the RL environment for CB is created on top of the SLS, following the format of Gymnasium [20]. For training, each episode consists of an independent SLS simulation run and includes a reset period followed by one RL step. The reset duration is 500 ms and the step duration is 1500 ms. The decay exponent of $\epsilon$ for the $\epsilon$-greedy algorithm is ~50, and the learning rate of the Adam algorithm is $10^{-3}$. The other algorithm parameters all use the example values in Section III.A. After training, to evaluate the performances of the agents in more practical scenarios when they are deployed for continuous inferencing, we use longer inferencing episodes, each of which includes a reset period followed by 10 RL steps. The reset duration is 500 ms and the step duration is 1000 ms, and so each inferencing episode consists of 10.5 s of SLS simulation (10 s for inferencing).

For each choice of reward function, we train the DRL agent with 500 independent episodes and the observed mean Q-values seem to have converged (except for the QoS-threshold reward, which is unstable and diverges), as shown in Fig. 5. Then the trained models are evaluated using 10 inferencing episodes, together with the baseline configuration when no cell DTX/DRX is used. Since each episode consists of 10 s of inferencing simulation on an independent deployment of 21 cells, in total we have 210 cells.[3] For the reward functions $r_{lin}$ and $r_{QoS\_approx}$, we try different values of their respective coefficient/constants and compare the power consumptions and QoS performances of the trained agents. The best coefficient/constants we found from these comparisons are $c = 0.75$ for the linear reward, and $y_0 = 0.9$, $m = 2$, and $\alpha = 3$ for the approximated QoS reward.

To demonstrate the performances of these agents in different traffic load scenarios, we divide the 210 cells into three categories based on their respective average PRB utilizations when no cell DTX/DRX is used (i.e., the baseline): 1) light traffic load, less than 40% PRB utilization; 2) medium traffic load, 40% ~ 80% PRB utilization; and 3) heavy traffic load, more than 80% PRB utilization. In Fig. 6 and Fig. 7 we respectively compare the average power consumptions and achieved data rates of the trained AI agents, using either the reward function $r_{lin}$ or $r_{QoS\_approx}$ (with the constants above), together with the baseline. We can see that compared to the baseline, both agents can save a considerable amount of energy in light and medium traffic scenarios ($\geq 45\%$ and $\geq 22\%$, respectively). In the heavy traffic scenario, while the agent with linear reward still provides some energy saving (∼12%), the reward $r_{QoS\_approx}$ leads to a slightly increased power consumption (∼2%, most likely due to increased ICI[4]). In terms of the achieved data rates, the approximate QoS reward provides a superior QoS performance: it keeps the data rate loss very small ($\leq 1\%$) in all scenarios. The linear reward, however, sacrifices a bit more QoS (up to 3% data rate degradation), especially in the heavy traffic scenario, in exchange of more energy saving. Hence if the QoS requirement is stringent, then the reward $r_{QoS\_approx}$ is recommended as it is more resilient to traffic load changes and can sacrifice some energy saving (e.g., in heavy traffic case) to keep the QoS loss small.

---
[3] The performance data of these 210 cells seem to be statistically representative enough for our performance evaluation purpose. The main limitation of using more inferencing episodes is the SLS simulation time.

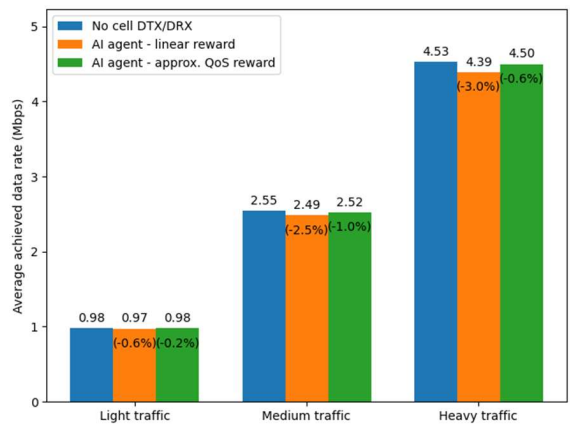

Fig. 7. Achieved data rates of AI agents and baseline.

## V. Conclusions and Future Work

In this paper we study the configuration optimization problem for 3GPP Rel-18 cell DTX/DRX to save maximum energy possible without hurting the QoS. By leveraging DRL architecture and the CB model, together with our novel nonlinear reward function design, we are able to train an AI agent that always tries to select the best possible cell DTX/DRX configuration under any network and traffic conditions. Our simulation results show that the agent can save up to ~45% energy depending on the traffic load, while always maintaining no more than ~1% degradation of the achieved data rate on average, compared to the case without cell DTX/DRX.

One future research direction is to incorporate the start offset parameter into the RL design. As mentioned earlier, this parameter is set to random in inference mode to reduce ICI. Instead of interference avoidance, however, if we can jointly design the start offset of the cells to achieve interference coordination, then the system performance may further improve. This joint design requires exchange of information between neighboring cells during learning, and for which we are currently considering to leverage the graph neural network (GNN) architecture (see, e.g., [21]). Another future direction is to study how to integrate the L1 group signaling DCI format 2_9 [15] into our current design, when urgent delay-sensitive transmission is needed. Simple reactive schemes that use L1 signaling to disable cell DTX/DRX whenever needed can be studied first, and further improvement using more sophisticated methods such as RL can be considered as the next step.

---

[4] Compared to the case without cell DTX/DRX, ICI may increase in a heavied loaded cell (hence always active) during the DTX active period of a neighboring cell. Hence in this time to transmit the same amount of data more power may be needed (e.g., because of lower MCS) at the cell.

References

[1] NGMN Alliance, "Green Future Networks: Network Energy Efficiency," whitepaper, Dec. 2021.

[2] ONF, "SMaRT-5G Project-Sustainable Mobile and RAN Transformation (SMaRT)," Open Networking Foundation, White Paper, 2023.

[3] 3GPP, TR 38.864, "Study on Network Energy Savings," 2022.

[4] NGMN Alliance, "Network Energy Efficiency Phase 2," whitepaper, Oct. 2023.

[5] P. Frenger, P. Moberg, J. Malmodin, Y. Jading and I. Godor, "Reducing Energy Consumption in LTE with Cell DTX," in 2011 IEEE 73rd Vehicular Technology Conference (VTC Spring), 2011.

[6] K.-H. Lin, H.-H. Liu, K.-H. Hu, A. Huang and H.-Y. Wei, "A Survey on DRX Mechanism: Device Power Saving From LTE and 5G New Radio to 6G Communication Systems," IEEE Communications Surveys & Tutorials, vol. 25, no. 1, pp. 156-183, 2023.

[7] A. De Domenico, V. Savin, D. Ktenas and A. Maeder, "Backhaul-aware small cell DTX based on fuzzy Q-Learning in heterogeneous cellular networks," in 2016 IEEE International Conference on Communications (ICC), 2016.

[8] A. De Domenico and D. Kténas, "Reinforcement learning for interference-aware cell DTX in heterogeneous networks," in 2018 IEEE Wireless Communications and Networking Conference (WCNC), 2018.

[9] R. S. Sutton and A. G. Barto, Reinforcement Learning: An Introduction, 2nd ed., The MIT Press, 2018.

[10] P. Bruhn and G. Bassi, "Machine Learning Based C-DRX Configuration Optimization for 5G," in Mobile Communication - Technologies and Applications; 25th ITG-Symposium, 2021.

[11] K. Boutiba and A. Ksentini, "On using Deep Reinforcement Learning to balance Power Consumption and Latency in 5G NR," in ICC 2023 - IEEE International Conference on Communications, 2023.

[12] 3GPP, TS 38.331, "Radio Resource Control (RRC) protocol specifications," 3GPP, Technical Specification Group Radio Access Network, NR, 2024.

[13] O-RAN Alliance, "O-RAN: Towards an Open and Smart RAN," whitepaper, Oct. 2018.

[14] CATT-R1-2211210, "Network energy saving techniques in time, frequency and spatial domain," 3GPP TSG RAN WG1#111, Toulouse, France, 2022.

[15] 3GPP, TS 38.212, "Multiplexing and channel coding," 3GPP Technical Specification Group Radio Access Network, NR, 2024.

[16] T. Islam, D. Lee and S. S. Lim, "Enabling Network Power Savings in 5G-Advanced and Beyond," IEEE Journal on Selected Areas in Communications, vol. Vol.41, no. No.6, pp. 1888-1899, June 2023.

[17] A. Paszke and M. Towers, "PyTorch Reinforcement Learning (DQN) Tutorial," [Online]. Available: https://docs.pytorch.org/tutorials/intermediate/reinforcement_q_learning.html.

[18] D. P. Kingma and J. Ba, "Adam: A Method for Stochastic Optimization," in International Conference on Learning Representations (ICLR), 2015.

[19] 3GPP, TR 36.889, "Study on Licensed-Assisted Access to Unlicensed Spectrum," 2015.

[20] Farama Foundation, "Gymnasium Documentation," [Online]. Available: https://gymnasium.farama.org/.

[21] O. Semiari, H. Nikopour and S. Talwar, "Graph Reinforcement Learning for QoS-Aware Load Balancing in Open Radio Access Networks," 2025. [Online]. Available: https://arxiv.org/abs/2504.19499.